\documentclass[]{tLCT2e}

\usepackage{amsmath,amssymb,amsthm}
\usepackage{graphicx}
\usepackage{subcaption}
\usepackage{verbatim}
\usepackage{hyperref}
\usepackage{bm}
\usepackage{color}
\usepackage{natbib}

\newcommand{\tr}{\operatorname{tr}}

\providecommand{\U}[1]{\protect\rule{.1in}{.1in}}

\begin{document}
	\title{Contributions of Repulsive and Attractive Interactions to Nematic Order}
\author{Peter Palffy-Muhoray$^1$\thanks{$^1$Corresponding author. Email: mpalffy@kent.edu} \\
		{\em{Liquid Crystal Institute, Kent State University, OH, USA}}\\
		\vspace{6pt}
Jamie M. Taylor$^2$\thanks{$^2$Email: jtayl139@kent.edu} \\
			{\em{Liquid Crystal Institute and Department of Mathematical Sciences, Kent State University, OH, USA}}\\
			\vspace{6pt}
Epifanio G. Virga$^3$\thanks{$^3$Email: eg.virga@unipv.it} \\
		{\em{Dipartimento di Matematica, Universit\`a di Pavia, Via Ferrata 5, I-27100 Pavia, Italy}}\\
		\vspace{6pt}
Xiaoyu Zheng$^4$\thanks{$^4$Email: zheng@math.kent.edu} \\
		{\em{Department of Mathematical Sciences, Kent State University, OH, USA}}\\		
		\received{Received 00 Month 20XX; final version received 00 Month 20XX} }
	\maketitle
\begin{abstract}
Both repulsive and attractive molecular interactions can be used to explain the onset of nematic order. The object of this paper is to combine these two nematogenic molecular interactions in a unified theory. This attempt is not unprecedented; what is perhaps new is the focus on the understanding of nematics in the high density limit. There, the orientational probability distribution is shown to exhibit a unique feature: it has compact support on configuration space. As attractive interactions are turned on, the behavior changes, and at a critical attractive interaction strength, thermotropic behavior of the Maier-Saupe type is attained.
\end{abstract}


\section{Introduction}\label{sec:intro}
In spite of their great scientific appeal and tremendous practical usefulness,
liquid crystals remain incompletely understood. Nematics are typically
described either in terms of attractive interactions, as in Maier-Saupe
theory~\cite{MaierSaupe58}, or in terms of repulsive interactions, as in
Onsager theory~\cite{Onsager49}. Attractive interactions are based on
long-range London dispersion, while repulsive interactions are based on
short-range steric effects. Although both interactions depend on particle
position and orientation, they are fundamentally different, and in some sense,
complementary. In the hard particle limit, the potential of steric interaction
is zero when the particles do not interpenetrate and infinite when they do not, while
the potential of the attractive interaction is zero when the particles
interpenetrate, and nonzero when they do not. One may regard position as the dominant
coordinate in steric interactions, with orientation playing a less important
role, while the opposite seems to hold in the case of attractive interactions. Position and orientation
are distinguished by the fact that two particles can have the same
orientation, whereas (the centers of mass of) two (convex) particles cannot have
the same position, reminiscent of fermions and bosons. This fundamental
difference may explain the great challenges that remain in the statistical
mechanics of rigid bodies, such as hard spheres with steric interactions, and
the great successes of relatively straightforward orientation-based models,
such as the Weiss theory of ferromagnetism or the Maier-Saupe theory of
nematics with long range attractive interactions.

Although it is expected that in common thermotropic nematics both short range
repulsive and long range attractive interactions play a role, only a few
models incorporate them consistently.\footnote{The reader is referred to the closing Sect.~\ref{sec:summary} for a short discussion on  pre-existing models; here, it would disrupt the flow of our presentation and overshadow the novelty of our approach.} We have recently become interested in
understanding the behavior of dense nematics. Of particular interest is the
high density regime, when the system begins to run out of available phase
space, and the  pressure diverges as the density approaches
a critical value, while the  free energy itself may remain bounded.

After having established on a rigorous basis the limit of dilution under which the Onsager theory is justifiable \cite{palffy-muhoray:missing}, we continue our inquiries by considering dense systems where hard ellipsoids
interact via purely steric excluded volume effects. Using a mean-field
approach, we have been able to describe the behavior, and calculate the order
parameter as function of density~\cite{PRE17}. An unusual result of our model
is that the single particle orientational distribution function has compact
support; that is, the probability of some range of particle orientations is
strictly zero. Predictions of our model for the particle volume fraction at the
nematic-isotropic transition as function of aspect ratios are in good
agreement with existing results from molecular dynamics simulations.

In this paper, we first review results from earlier work on attractive and repulsive interactions in nematics in the low density limit (Sect.~\ref{sec:low_density}). We then
consider the high density limit, where we consider the effects of attractive interactions in addition to hard core repulsion (Sect.~\ref{sec:high_density}). Section~\ref{sec:uniaxiality} is where we introduce our simplifying assumption about the uniaxiality of the ordered phases we are investigating. In Sect.~\ref{sec:results} we illustrate our numerical results and dwell  on their physical interpretation. Finally, in the closing Sect.~\ref{sec:summary}, we give our perspective on our results in view of both old theories and new challenges ahead.

With the
consistent inclusion of both repulsive and attractive effects in the model, we
hope to gain insights into the relative importance of these two dissimilar, but conspiring interactions.

\section{Attractive and repulsive interactions in the low density limit}
\label{sec:low_density}
As indicated in \cite{PCMI}, the Maier-Saupe and Onsager models are rather
similar. This can be clearly seen if the excluded volume in Onsager theory is
modified, and replaced by an approximation for the excluded volume for ellipsoids.

For identical hard ellipsoids of revolution with length $L$, width $W$ and
volume $v_{0}$, a simple approximation for the excluded volume is
\begin{align}
V_{exc}(\mathbf{\hat{l}}_{1},\mathbf{\hat{l}}_{2}) &  =C-\frac{2D}{3}\left(
\boldsymbol{\sigma}\mathbf{ (\hat{l}}_{1}):\boldsymbol{\sigma}\mathbf{(\hat
	{l}}_{2})\right)  ,\label{eq_excluded_vol}\\
C &  =v_{0}\frac{4}{3}(\frac{L}{W}+\frac{W}{L}+4),\label{eq_c}\\
D &  =v_{0}\frac{4}{3}(\frac{L}{W}+\frac{W}{L}-2),\label{eq_d}%
\end{align}
where $\boldsymbol{\sigma}\mathbf{(\hat{l})}=\frac{1}{2}(3\mathbf{\hat{l}%
	\hat{l}}-\mathbf{I)}$ denotes the orientation descriptor of a particle with
symmetry axis oriented along $\mathbf{\hat{l}}$, with the shape parameters
$D\geq0$ and $C>D$. With this expression for the excluded volume, the
Maier-Saupe and modified Onsager results are essentially equivalent in that
density in the Onsager model plays the role of inverse temperature. Results
for the combined Maier-Saupe and modified Onsager theory are shown in Fig.~\ref{fig-PCMI}
below. There, the order parameter $S$ at the critical points of the free
energy is shown as a function of dimensionless temperature $T/T^{\prime}$ and
density $\rho_{n}$.

\begin{figure}[th]
	\centering
	\includegraphics[width=.5\linewidth]{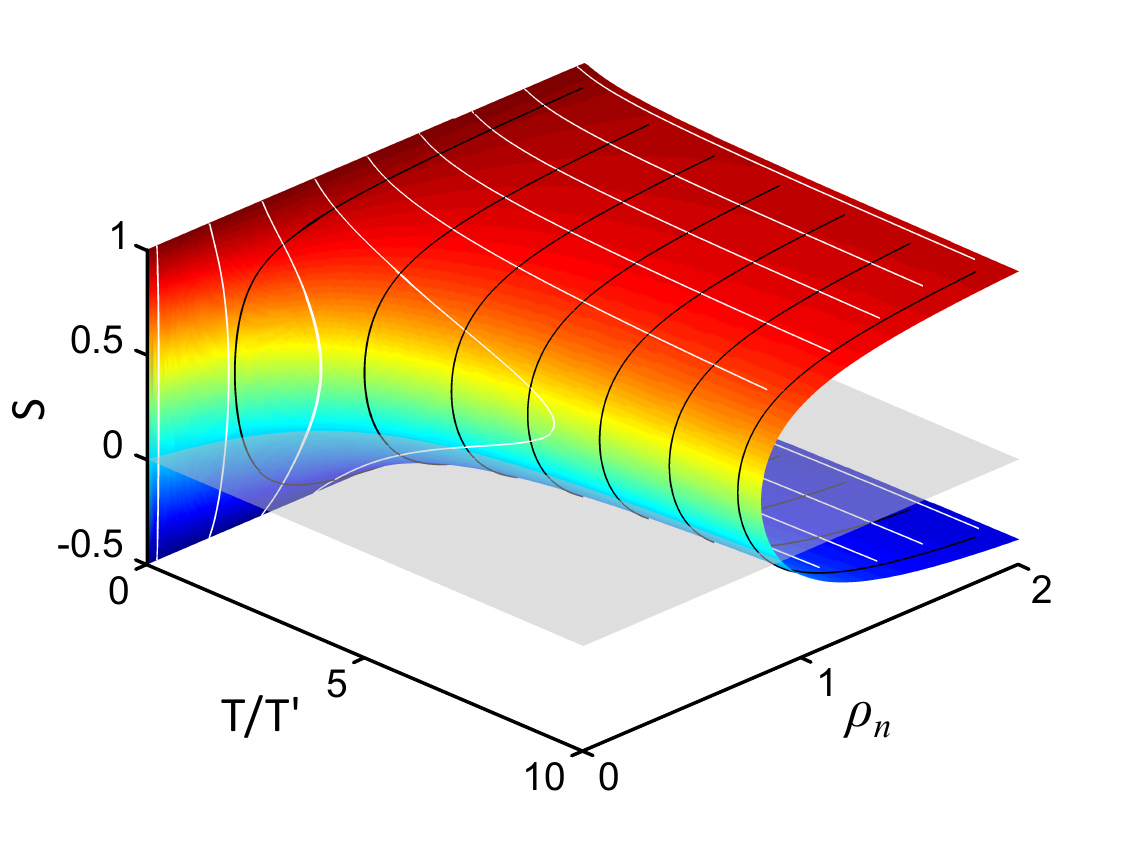}\caption{Equilibrium order parameter $S$ as a
		function of dimensionless temperature and density for the combined Maier-Saupe and modified Onsager theory~\cite{PCMI}. $T^{\prime}= \frac{U_{0}b}{k D}$ and $\rho_{n} = \frac{1}{5} \rho_{0} D$, where $U_0$ fixes the energy scale and $\rho_0$ is the number density (see \eqref{eq_attraction} and \eqref{eq:rho_0_definition}, respectively, for their formal definition).}
	\label{fig-PCMI}%
\end{figure}

We note that a much better approximation to the excluded volume of ellipsoids
is obtained for different parametrization of $C$ and $D$ in terms of aspect
ratio~\cite{PRE17}.

\section{Attractive and repulsive interactions in the high density limit}
\label{sec:high_density}
In this section, we look into the behavior of the system near the high
density limit. We begin by outlining the building blocks of our model.

We consider a one-component system consisting of hard ellipsoids interacting
pairwise via both attractive and repulsive potentials. The configurational
Helmholtz free energy of a system of $N$ particles, within an additive
constant, is%
\begin{equation}
F=-kT\ln\frac{1}{N!}\int_{\Omega^{N}}e^{-\frac{1}{kT}\sum_{1\leq i<j\leq
		N}U_{ij}}d\mathbf{q}_{1}\cdots d\mathbf{q}_{N},
\end{equation}
where $k$ is the Boltzmann constant, $\mathbf{q}_{i}$ is the generalized (orientational and positional)
coordinate of the $i^{th}$ particle, and $U_{ij}=U(\mathbf{q}_{i}%
,\mathbf{q}_{j})$ is the total interaction energy between particles $i$ and
$j$, the sum is over all pairs of particles.

The interaction potential consists of a long range attractive and a short
range repulsive part;%
\begin{equation}
U(\mathbf{q}_{i},\mathbf{q}_{j})=U^{A}(\mathbf{q}_{i},\mathbf{q}_{j}%
)+U^{R}(\mathbf{q}_{i},\mathbf{q}_{j}).
\end{equation}
The attractive interaction potential is of the form \cite{PCMI},
\begin{equation}\label{eq:dispersion_forces}
U^{A}(\mathbf{q}_{i},\mathbf{q}_{j})=-\frac{K}{r^{6}}(\boldsymbol{\alpha}%
_{i}:\boldsymbol{\alpha}_{j}+3\tr\boldsymbol{\alpha}_{i}\tr\boldsymbol{\alpha
}_{j}),
\end{equation}
where $\boldsymbol{\alpha} $ is a polarizability tensor, $\tr$ denotes the trace, $K>0$ is an interaction
constant, and $r$ is the separation between the centers of particles. The
repulsive interaction has the form
\begin{equation}
U^{R}(\mathbf{q}_{i},\mathbf{q}_{j})=\left\{
\begin{array}
[c]{l}%
\infty\text{, if particles interpenetrate,}\\
0,\text{ otherwise.}%
\end{array}
\right.
\end{equation}

We write
\begin{equation}
G_{N}=\int_{\Omega^{N}}e^{-\frac{1}{kT}\sum_{1\leq i<j\leq N}U_{ij}%
}d\mathbf{q}_{1}\cdots d\mathbf{q}_{N},
\end{equation}
where quantity $G_{N}$ can be thought of as the number of states available to
$N$ distinguishable particles.

Since all particles are equivalent, we write, as in \cite{PRE17}, in the
spirit of mean field theory,%
\begin{equation}
G_{N}=\left\langle \int_{\Omega}e^{-\frac{1}{kT}(\sum_{j=2}^{N}U_{1j}^{A}%
	+\sum_{j=2}^{N}U_{1j}^{R})}d\mathbf{q}_{1}\right\rangle ^{N},\label{at1}%
\end{equation}
where the average $\left\langle\cdot  \right\rangle $ is calculated with respect to the $(N-1)$-particle Gibbs
distribution. We further pursue our mean-field strategy, by replacing  the attractive part of the potential in
Eq.~(\ref{at1}) by the single particle pseudopotential,
\begin{equation}\label{eq:attractive_peseudopotential}
U^{A}(\mathbf{q}_{1})=\sum_{j=2}^{N}U_{1j}^{A}=\frac{1}{2}\int\rho
(\mathbf{q}_{2})U_{12}^{A}(\mathbf{q}_{1},\mathbf{q}_{2})d\mathbf{q}_{2},
\end{equation}
where $\rho(\mathbf{q}_{2})$ is the number density of particles with
generalized coordinate $\mathbf{q}_{2}$. According to common folklore, the factor $\frac12$ has been introduced in \eqref{eq:attractive_peseudopotential}to avoid ``double counting'' of particles; more precisely, its role is keeping the total attractive energy unchanged in passing from the discrete to a continuum representation:
\begin{equation}\label{eq:keeping_energy}
\frac12\sum_{i\neq j=1}^NU^A_{ij}=\int_{\Omega}\rho(\mathbf{q}_{1})U^A(\mathbf{q}_{1})d\mathbf{q}_{1}.
\end{equation}
We thus rewrite \eqref{at1} as
\begin{equation}
G_{N}=\left\langle \int_{\Omega}e^{-\frac{1}{kT}[U^{A}(\mathbf{q}_{1}%
	)+\sum_{j=2}^{N}U_{1j}^{R}]}d\mathbf{q}_{1}\right\rangle ^{N},
\end{equation}
where, despite $U^A$ being now a singe-particle function, the average $\left\langle\cdot\right\rangle$ is still computed with respect to the $(N-1)$-particle Gibbs distribution. A further step in the way of simplifying our theory is posing
\begin{equation}
G_{N}=\left(  \int_{\Omega}e^{-\frac{1}{kT}U^{A}(\mathbf{q}_{1})}%
[1-W(\mathbf{q}_{1})]d\mathbf{q}_{1}\right) ^{N},
\end{equation}
where
\begin{equation}
W(\mathbf{q}_{1})=\left\langle 1-e^{-\frac{1}{kT}\Sigma_{j=2}^{N}U_{1j}^{R}%
}\right\rangle
\end{equation}
is the average excluded volume fraction. We estimate this quantity as
\begin{equation}
W(\mathbf{q}_{1})=\lambda\int_{\Omega}\rho(\mathbf{q}_{2})\left[1-e^{-\frac
	{U^{R}(\mathbf{q}_{1},\mathbf{q}_{2})}{kT}}\right]d\mathbf{q}_{2},
\end{equation}
where, as shown in  \cite{PRE17}, $\lambda$ has the role of a dimensionless  phenomenological parameter.\footnote{which might better be thought of as phenomenological function of the number density $\rho_0$ introduced shortly below, in \eqref{eq:rho_0_definition}.}

The free energy then becomes
\begin{equation}
F=-kT\ln\frac{1}{N!}\left(  \int_{\Omega}e^{-\frac{1}{kT}U^{A}(\mathbf{q}%
	_{1}\mathbf{)}}[1-W(\mathbf{q}_{1})]d\mathbf{q}_{1}\right)  ^{N},\label{2}%
\end{equation}
which is the cornerstone of our model.
The integral in Eq.~(\ref{2}) can be regarded as the weighted fraction of the
total volume available to particle $1$, or the weighted free volume fraction.

This can be written in the density functional form, using the procedure we
developed previously as described in \cite{PCMI, PRE17},
\begin{align}
F  & =kT\int_{\Omega}\rho(\mathbf{q}_{1})\ln\rho(\mathbf{q}_{1})d\mathbf{q}%
_{1}\mathbf{+}\int_{\Omega}\rho(\mathbf{q}_{1}\mathbf{)}U^{A}(\mathbf{q}%
_{1}\mathbf{)}d\mathbf{q_{1}}\nonumber\\
& -kT\int_{\Omega}\rho(\mathbf{q}_{1})\ln\left(1-\lambda\int_{\Omega}%
\rho(\mathbf{q}_{2})\left[1-e^{-\frac{1}{kT}U^{R}(\mathbf{q}_{1},\mathbf{q}_{2})}\right]d\mathbf{q}_{2}\right)  d\mathbf{q}_{1}.
\end{align}
It is convenient to write the generalized coordinates in terms of position and
orientation, then
\begin{align}
\label{eq_energy_1}F &  =kT\int_{\mathbb{S}^{2}}\int_{\mathcal{B}}%
\rho(\mathbf{r}_{1},\mathbf{\hat{l}}_{1})\ln\rho(\mathbf{r}_{1},\mathbf{\hat
	{l}}_{1})d^{3}\mathbf{r}_{1}d^{2}\mathbf{\hat{l}}_{1}\mathbf{+}\int_{\Omega
}\rho(\mathbf{r}_{1},\mathbf{\hat{l}}_{1}\mathbf{)}U^{A}(\mathbf{r}%
_{1},\mathbf{\hat{l}}_{1}\mathbf{)}d^{3}\mathbf{r}_{1}d^{2}\mathbf{\hat{l}%
}_{1}\nonumber\\
&  -kT\int_{\mathbb{S}^{2}}\int_{\mathcal{B}}\rho(\mathbf{r}_{1}%
,\mathbf{\hat{l}}_{1})\ln\left(  1-\lambda\int_{\mathbb{S}^{2}}\int
_{\mathcal{B}}\rho(\mathbf{r}_{2},\mathbf{\hat{l}}_{2})\left[1-e^{-\frac{1}{kT}
	U^{R}(\mathbf{r}_{1},\mathbf{\hat{l}}_{1},\mathbf{r}_{2},\mathbf{\hat{l}}%
		_{2})}\right]d^{3}\mathbf{r}_{2}d^{2}\mathbf{\hat{l}}_{2}\right)
d^{3}\mathbf{r}_{1}d^{2}\mathbf{\hat{l}}_{1},
\end{align}
where $\rho(\mathbf{r},\mathbf{\hat{l}})$ is the number density of particles at
$\mathbf{r}$ with orientation
$\mathbf{\hat{l}}$. $\mathbb{S}^{2}$ denotes orientation space, the
surface of the unit sphere. $\mathcal{B}$ denotes position space, occupied by
particles. In a homogeneous system, the density is independent of $\mathbf{r}%
$, so we write
\begin{equation}\label{eq:rho_0_definition}
\rho(\mathbf{r},\mathbf{\hat{l}})=\rho_{0}f(\mathbf{\hat{l}}),
\end{equation}
where $\rho_{0}$ now is simply the number density of particles, and
$f(\mathbf{\hat{l}})$ is the single particle orientational distribution
function satisfying%
\begin{equation}\label{eq:normalization_condition}
\int_{\mathbb{S}^{2}}f(\mathbf{\hat{l}})d^{2}\mathbf{\hat{l}}=1.
\end{equation}
Integrating Eq.~(\ref{eq_energy_1}) over $\mathbf{r}_{1}$ and $\mathbf{r}_{2}$
gives
\begin{align}
F  & =kT\rho_{0}V\left\{ \ln\rho_{0}+\int_{\mathbb{S}^{2}}f(\mathbf{\hat{l}%
}_{1})\ln f(\mathbf{\hat{l}}_{1})d^{2}\mathbf{\hat{l}}_{1}\mathbf{+}\frac
{1}{kT}\int_{\mathbb{S}^{2}}f(\mathbf{\hat{l}}_{1})U^{A}%
(\mathbf{\hat{l}}_{1})d^{2}\mathbf{\hat{l}}_{1} \right. \nonumber\\
& \left. -\int_{\mathbb{S}^{2}}f(\mathbf{\hat{l}}_{1})\ln\left(1-\lambda\rho
_{0}\int_{\mathbb{S}^{2}}f(\mathbf{\hat{l}}_{2})V_{exc}(\mathbf{\hat{l}}%
_{1},\mathbf{\hat{l}}_{2})d^{2}\mathbf{\hat{l}}_{2}\right)d^{2}\mathbf{\hat{l}}%
_{1}\right\} ,\label{eq_energy}%
\end{align}
where
\begin{equation}
V_{exc}(\mathbf{\hat{l}}_{1},\mathbf{\hat{l}}_{2})=\int_{\mathcal{B}%
}\left[1-e^{-\frac{1}{kT}U^{R}(\mathbf{r}_{1},\mathbf{l}_{1},\mathbf{r}_{2}%
		,\mathbf{l}_{2})}\right]d^{3}\mathbf{r}%
\end{equation}
is  excluded volume of two particles (integrated in their relative position $\mathbf{r}=\mathbf{r}_2
	-\mathbf{r}_1$).

If $\rho_{0}$ is small, one can expand the logarithm and recover the theory in
the dilute limit \cite{PCMI}. We note that the expansion is only valid when
\begin{equation}
\lambda\rho_{0}\int_{\mathbb{S}^{2}}f(\mathbf{\hat{l}}_{2})V_{exc}%
(\mathbf{\hat{l}}_{1},\mathbf{\hat{l}}_{2})d^{2}\mathbf{\hat{l}}_{2}<1,
\end{equation}
since the argument of logarithm must be positive.

To capture accurately the phase behavior, we keep the full logarithmic
dependence in this work; this is the salient feature of our approach. This
form of the free energy is the origin of a remarkable phenomenon: without the
attractive interaction, above a critical value of $\rho_{0}$, the equilibrium
distribution function $f(\mathbf{\hat{l}})$ is zero on a region of the
orientation space with positive measure; that is, at high densities, some
regions of orientation space are not accessible to particles.

Since London dispersion is a function of the particle polarizability tensor
$\boldsymbol{\alpha}$, which is linear in the orientation descriptor
$\boldsymbol{\sigma}(\mathbf{l})$, we can write the attractive
interaction explicitly as follows \cite{PCMI},%
\begin{align}
\label{eq_attraction}U^{A}(\mathbf{l}_{1}) &  =\frac{1}{2}\int\rho
(\mathbf{r}_{2},\mathbf{\hat{l}}_{2})U^{A}(\mathbf{r}_{1},\mathbf{\hat
	{l}}_{1},\mathbf{r}_{2},\mathbf{\hat{l}}_{2})d^{3}\mathbf{r}_{2}%
d\mathbf{\hat{l}}_{2}\nonumber\\
&  =-\frac{1}{2}\rho_{0}U_{0}\left(a+b\frac{2}{3}\boldsymbol{\sigma} \mathbf{(\hat
	{l}}_{1}):\mathbf{Q}\right),%
\end{align}
where $U_{0}$ is an energy (appropriately related to $K$ in \eqref{eq:dispersion_forces}), and $a$ and $b$, the isotropic and anisotropic
parameters of the potential, have units of volume.

Substitution of the attractive interaction (\ref{eq_attraction}) and the
excluded volume (\ref{eq_excluded_vol}) gives the free energy density (per unit volume),
\begin{align}
\mathcal{F}=  & kT\rho_{0}\left\{ \ln\rho_{0}+\int_{\mathbb{S}^{2}%
}f(\mathbf{\hat{l}})\ln f(\mathbf{\hat{l}})d^{2}\mathbf{\hat{l}}-\frac{1}%
{2kT}\int_{\mathbb{S}^{2}}f(\mathbf{\hat{l}})\rho_{0}U_{0}\left(a+b\frac{2}{3}\boldsymbol{\sigma}\mathbf{ (\hat{l}):Q}\right)d^{2}\mathbf{\hat{l}}\right.
\nonumber\\
& \left.  -\int_{\mathbb{S}^{2}}f(\mathbf{\hat{l}})\ln\left[1-\rho_{0}\left(c-d\frac
{2}{3}\boldsymbol{\sigma}\mathbf{(\hat{l})}:\mathbf{Q}\right)\right]d^{2}\mathbf{\hat{l}}\right\},
\end{align}
where
\begin{equation}
\mathbf{Q}=\int_{\mathbb{S}^{2}}\boldsymbol{\sigma}\mathbf{(\hat{l}%
	)}f(\mathbf{\hat{l}})d^{2}\mathbf{\hat{l},}%
\end{equation}
is the symmetric and traceless tensor, which is the familiar orientational order descriptor of nematic liquid crystals. Here the
excluded volume constants are defined as $c=\lambda C$ and $d=\lambda D$. Note that if
$f(\mathbf{\hat{l}})$ is admissible, it must satisfy
\begin{equation}
1-\rho_{0}\left(c+d\frac{2}{3}\boldsymbol{\sigma} \mathbf{(\hat{l})}:\mathbf{Q}\right)>0,
\end{equation}
for all $\mathbf{\hat{l}}$ with $f(\mathbf{\hat{l}})>0$; that is, if the argument of the
logarithm is  \emph{negative}, then $f(\mathbf{\hat{l}})$ must be zero.

It is convenient to introduce an auxiliary dimensionless parameter as in \cite{PRE17},%
\begin{equation}
\phi=\frac{\rho_{0}c-1}{\rho_{0}d}\in(-\infty,1],\label{eq_phi}%
\end{equation}
which is an increasing function of number density $\rho_{0}=\frac{1}{c-d\phi}
$. In the very dilute limit, $\rho_{0}\rightarrow0^{+},\phi\rightarrow-\infty
$; in the dense packing limit, $\rho_{0}\rightarrow\left(\frac{1}{c-d}\right)^{-}%
,\phi\rightarrow1^{-}$, this corresponds to the densest packing fraction. It is remarkable that the information of the
particle shape is completely subsumed in the effective density $\phi$. Retaining only relevant
terms, we express the free energy density $\mathcal{F}$ in the following form
\begin{equation}\label{eq:pre_free_energy}
\mathcal{F}=kT\rho_{0}\int_{\mathbb{S}^{2}}\left[f(\mathbf{\hat{l}})\ln
f(\mathbf{\hat{l}})-\tau f(\mathbf{\hat{l}%
	)}\left(\frac{a}{b}+\frac{2}{3}\boldsymbol{\sigma}\mathbf{(\hat{l}):Q}\right)-f(\mathbf{\hat{l}%
})\ln\left(  \frac{2}{3}\boldsymbol{\sigma}(\mathbf{\hat{l}):Q}-\phi\right)\right]
d^{2}\mathbf{\hat{l}},
\end{equation}
where the dimensionless parameter,
\begin{equation}\label{eq:tau_definition}
\mathbf{\tau}=\frac{\rho_{0}U_{0}b}{kT},
\end{equation}
plays the role of an effective inverse temperature.
More explicitly, Eq.~\eqref{eq:pre_free_energy} can also be written as
\begin{align}
\label{eq_energy_2}\mathcal{F}= & kT\rho_{0}\int_{\mathbb{S}^{2}}\left[
f(\mathbf{\hat{l}}_{1})\ln f(\mathbf{\hat{l}}_{1})-\frac{1}{2}\tau
f(\mathbf{\hat{l}}_1)\left(\frac{a}{b}+\frac{2}{3}\boldsymbol{\sigma}(\mathbf{\hat{l}}_1):%
\int_{\mathbb{S}^{2}}\boldsymbol{\sigma}\mathbf{(\hat{l}}_{2}\mathbf{)}%
f(\mathbf{\hat{l}}_{2})d^{2}\mathbf{\hat{l}}_{2}\right)\right. \nonumber\\
&  \left.-f(\mathbf{\hat{l}}_{1})\ln\left(  \frac{2}{3}\boldsymbol{\sigma
}\mathbf{(\hat{l}}_{1}):\int_{\mathbb{S}^{2}}\boldsymbol{\sigma}%
\mathbf{(\hat{l}}_{2}\mathbf{)}f(\mathbf{\hat{l}}_{2})d^{2}\mathbf{\hat{l}%
}_{2}-\phi\right)\right]  d^{2}\mathbf{\hat{l}}_{1}.
\end{align}

We next minimize the free energy density (\ref{eq_energy_2}) with respect to
the orientational distribution function $f(\mathbf{\hat{l}})$. There are two
constraints. First, the distribution function is normalized to unity subject to $f(\mathbf{\hat{l}})\ge0$. Second, the argument of the logarithm, the free volume
fraction, must be positive where $f(\mathbf{\hat{l}})>0$; that is,
\begin{equation}
\frac{2}{3}\boldsymbol{\sigma}\mathbf{(\hat{l})}:\mathbf{Q}-\phi>0.
\end{equation}

Setting the first variation to zero and solving for the distribution function
gives the self-consistent equation for $f$ in terms of the order parameter tensor
$\mathbf{Q}$%
\begin{equation}
\label{eq_pdf}f(\mathbf{\hat{l}}_{1})=\left\{
\begin{array}
[c]{cc}%
\displaystyle\frac{\left(\frac{2}{3}\boldsymbol{\sigma}\mathbf{(\hat{l}}%
	_{1}\mathbf{)}:\mathbf{Q}-\phi\right)e^{\boldsymbol{\sigma}(\mathbf{\hat{l}}%
		_{1}):\left( \tau \frac{2}{3}\mathbf{Q}+\displaystyle\int_{\mathbb{S}^{2}%
		}f(\mathbf{\hat{l}}_{2})\frac{\frac{2}{3}\boldsymbol{\sigma}\mathbf{(\hat{l}%
			}_{2})}{\frac{2}{3}\boldsymbol{\sigma}(\mathbf{\hat{l}}_{2}):\mathbf{Q}-\phi
		}d^{2}\mathbf{\hat{l}}_{2}\right) }}{\displaystyle\int_{\mathbb{S}^{2}}%
	\left(\frac{2}{3}\boldsymbol{\sigma}\mathbf{(\hat{l}}_{1}\mathbf{)}:\mathbf{Q}%
	-\phi\right)e^{\boldsymbol{\sigma}(\mathbf{\hat{l}}_{1}):\left( \tau \frac{2}%
		{3}\mathbf{Q}+\displaystyle\int_{\mathbb{S}^{2}}f(\mathbf{\hat{l}}_{2}%
		)\frac{\frac{2}{3}\boldsymbol{\sigma}\mathbf{(\hat{l}}_{2})}{\frac{2}%
			{3}\boldsymbol{\sigma}(\mathbf{\hat{l}}_{2}):\mathbf{Q}-\phi}d^{2}%
		\mathbf{\hat{l}}_{2}\right) }} & \text{if }\frac{2}{3}\boldsymbol{\sigma
}\mathbf{ (\hat{l}):Q}-\phi>0,\\
0 & \text{otherwise}.
\end{array}
\right.
\end{equation}

We define an auxiliary order parameter tensor $\mathbf{\Psi}$ as
\begin{equation}
\mathbf{\Psi}=\tau \frac{2}{3}\mathbf{Q}+\int_{\mathbb{S}^{2}}f(\mathbf{\hat
	{l}})\frac{\frac{2}{3}\boldsymbol{\sigma}\mathbf{(\hat{l}})}{\frac{2}%
	{3}\boldsymbol{\sigma}\mathbf{(\hat{l})}:\mathbf{Q}-\phi}d^{2}\mathbf{\hat
	{l}.}%
\end{equation}
Finally, the equation for the orientational distribution function becomes
\begin{equation}
f(\mathbf{\hat{l}})=\frac{\left(\frac{2}{3}\boldsymbol{\sigma}\mathbf{(\hat
		{l}):Q}-\phi\right)e^{\boldsymbol{\sigma}\mathbf{(\hat{l}):\Psi}}}
{\displaystyle\int_{\mathbb{S}^{2}}\left(\frac{2}{3}\boldsymbol{\sigma}\mathbf{(\hat{l}):Q}-\phi
	\right)e^{\boldsymbol{\sigma}\mathbf{(\hat{l}):\Psi}d^{2}\mathbf{\hat{l}}}},
\end{equation}
where%
\begin{subequations}
\begin{align}
\mathbf{Q}&=\left\langle \boldsymbol{\sigma} \mathbf{(\hat{l})}\right\rangle ,\label{eq:Q_definition}\\
\mathbf{\Psi}&=\tau \frac{2}{3}\mathbf{Q}+\left\langle \frac{\frac{2}%
	{3}\boldsymbol{\sigma}(\mathbf{\hat{l}})}{\frac{2}{3} \boldsymbol{\sigma
	}\mathbf{(\hat{l})}:\mathbf{Q}-\phi}\right\rangle,\label{eq:Psi_definition}
\end{align}
\end{subequations}
with both averages $\left\langle\cdot\right\rangle$ now computed with respect to $f$ itself, as is customary in any mean-field approach.
The requirement of nonnegativity of the free volume fraction, $\frac{2}%
{3}\mathbf{\sigma(\hat{l})}:\mathbf{Q}-\phi>0$, results in the orientational
distribution function being zero in some regions of orientation space in the
absence of attractive interactions \cite{PRE17}.

The singular nature of the model makes suspect the use of the first variation
to obtain the Euler-Lagrange equation (\ref{eq_pdf}), as it is unclear whether the
free energy density is sufficiently differentiable for variations to be taken,
where the orientation distribution function is zero, and whether
finite-energy configurations even exist. However, by means of a duality argument, these analytical issues of models such as ours have been rigorously
addressed previously \cite{Jamie}. In fact, the auxiliary order parameter tensor
$\mathbf{\Psi}$ arises as, in a sense, a dual to the order parameter tensor $\mathbf{Q}$,
though its physical interpretation still remains
elusive. The analysis proves rigorously that the formal Euler-Lagrange
equation (\ref{eq_pdf}) is indeed satisfied by equilibria for $\phi<1$, while
for $\phi\ge1$ no finite energy configurations exist, which we interpret to be
the saturation limit of the model. Similar questions of differentiability can
be asked of the derivation of the equation of state (see
Eq.~\eqref{ref:eq_pressure} below), however our formal calculation presented in this
work can similarly be shown to be valid.

To derive the equation of state of our system of hard ellipsoids, we start
with the relation between the pressure and the free energy density,
\begin{equation}
P=-\mathcal{F+}\rho_{0}\frac{\partial\mathcal{F}}{\partial\rho_{0}},
\end{equation}
which gives
\begin{equation}
P+\frac{1}{2}\rho_{0}^{2}U_{0}\left( a+b\frac{2}{3}\mathbf{Q}^{2}\right)
=\rho_{0}kT\left( \frac{1}{1-\rho_{0}c}\right) \left\langle \frac{1}%
{1-\frac{2}{3}\phi^{-1}\mathbf{Q}:\mathbf{\sigma(\hat{l})}}\right\rangle
.\label{ref:eq_pressure}%
\end{equation}
The denominator of the second term on the right hand side of Eq.~\eqref{ref:eq_pressure}is the isotropic part of the excluded volume fraction, while the third term, the anisotropic contribution, is the reciprocal of the harmonic mean of the excluded volume fraction.

If $\mathbf{Q}=\bm{0}$, then we have
\begin{equation}
P+\frac{1}{2}\rho_{0}^{2}U_{0}a=\frac{\rho_{0}kT}{1-\rho_{0}c},
\end{equation}
as in the Van der Waals case. In general, Eq.~(\ref{ref:eq_pressure}) is our
equation of state. In high density regime, when $\phi$ approaches $1$, the
pressure approaches infinity.

\section{The assumption of uniaxiality}
\label{sec:uniaxiality}
Without external fields, the classical models, Maier-Saupe theory 
\cite{MaierSaupe58} for attractive interactions and Onsager theory 
\cite{Onsager49} for steric interactions, predict only uniaxial nematic order
at high densities \cite{Fatkullin05}. Here we also assume that $\mathbf{Q}$ is
uniaxial to simplify the presentation. There might also be biaxial
equilibrium solutions, like those found for the fully repulsive case in \cite{PRE17}; there,   biaxial solutions were
saddle points of the free energy. We expect that biaxial solutions, if they exist, would not be
stable when attractive interactions are included. We write%
\begin{equation}
\mathbf{Q}=\frac{S}{2}(3\mathbf{\hat{n}\hat{n}-I}),
\end{equation}
where $S$ is the scalar order parameter and $\mathbf{\hat{n}}$ is the nematic
director. One can show then that $\mathbf{\Psi}$ is also uniaxial, and shares the
same eigenframe with $\mathbf{Q}$ \cite{Jamie}, thus can be written as
\begin{equation}
\mathbf{\Psi}=\Psi(\mathbf{\hat{n}\hat{n}}-\frac{1}{3}\mathbf{I}).
\end{equation}
Then we have
\begin{align}
\mathbf{Q} &  :\boldsymbol{\sigma}(\mathbf{\hat{l}})\mathbf{=}\frac{3S}%
{2}P_{2}(\mathbf{\hat{n}}\cdot\mathbf{\hat{l}}),\nonumber\\
\mathbf{\Psi} &  :\boldsymbol{\sigma}(\mathbf{\hat{l}})=\Psi P_{2}%
(\mathbf{\hat{n}}\cdot\mathbf{\hat{l}}),
\end{align}
where $P_{2}(x)$ is the second order Legendre polynomial. The free energy
density becomes
\begin{equation}
\mathcal{F}=2\pi\rho_{0}kT\left(  \Psi S-\frac{1}{2}\tau(\frac{a}{b}+S^{2})-\ln\int_{\mathbb{S}_{+}^{2}}(SP_{2}(x)-\phi)e^{\Psi P_{2}%
	(x)}dx\right),
\end{equation}
where use of \eqref{eq:tau_definition} has been made,
$x=\mathbf{\hat{n}\cdot\hat{l}}=\cos\theta$, and $\mathbb{S}_{+}%
^{2}=\{x:SP_{2}(x)-\phi>0\}$. The uniaxiality assumption makes the analysis
and numerics more tractable. The distribution function is now given by%
\begin{equation}
f(x)=\left\{
\begin{array}
[c]{cc}%
\frac{\left(  SP_{2}(x)-\phi\right)  \exp(\Psi P_{2}(x))}{\int_{\mathbb{S}%
		_{+}^{2}}\left(  SP_{2}(x)\mathbf{-}\phi\right)  \exp(\Psi P_{2}(x))dx} &
\text{if }SP_{2}(x)-\phi>0,\\
0 & \text{otherwise}.
\end{array}
\right.  \text{ }%
\end{equation}
Instead of solving the above self-consistency equation for $f(x)$, we solve the
coupled equations for $\Psi$ and $S$,
\begin{subequations}\label{eq:self_consistency}
\begin{equation}
S=\frac{\int_{\mathbb{S}_{+}^{2}}P_{2}(x)\left(  SP_{2}(x)-\phi\right)
	\exp(\Psi P_{2}(x))dx}{\int_{\mathbb{S}_{+}^{2}}\left(  SP_{2}(x)\mathbf{-}%
	\phi\right)  \exp(\Psi P_{2}(x))dx}=\left\langle P_{2}(x)\right\rangle
,\label{eq_S_uni}%
\end{equation}%
\begin{equation}
\Psi=\tau S+\frac{\int_{\mathbb{S}_{+}^{2}}P_{2}(x)\exp(\Psi P_{2}%
	(x))dx}{\int_{\mathbb{S}_{+}^{2}}\left(  SP_{2}(x)-\phi\right)  \exp(\Psi
	P_{2}(x))dx}=\tau S+\left\langle \frac{P_{2}(x)}{SP_{2}(x)-\phi}\right\rangle
.\label{eq_lambda_uni}%
\end{equation}
\end{subequations}
Explicitly, the integration limits are
\begin{equation}
(x_{0},1),\text{ if }S   >0\quad\text{and}\quad
(0,x_{0}),\text{ if }S   <0,
\end{equation}
and $x_{0}>0$ is such that $P_{2}(x_{0})=\phi/S$.\footnote{We study $f$ only
	for $0\leq x\leq1$, as it is clearly even in $-1\leq x\leq1$.}

For a given effective density $\phi$ and effective inverse temperature $\tau$,
Eqs.~(\ref{eq_S_uni}) and (\ref{eq_lambda_uni}) can be solved
numerically for the scalar order parameters $S$ and $\Psi$, as well as for the cutoff
parameter $x_{0}$; the details of the algorithm will given elsewhere. Once
these are determined, the behavior of the system is known. We present the
results in the following section.

\section{Results}
\label{sec:results}
The behavior of the system without attractive interactions has been detailed
in \cite{PRE17}. The equilibrium uniaxial order parameter $S$ versus effective density $\phi$ is shown
in Fig.~\ref{fig-PRE} below. The stability of different branches is determined by
examining the local convexity of the free energy density. For $\phi<\phi
_{NI}\approx -0.224$, the system is in the isotropic state with $S=0$; at $\phi
=\phi_{NI}$, the system undergoes a first order transition to the nematic
state, with order parameter $S_{NI}\approx 0.545$. As $\phi\rightarrow 1$, the order
parameter $S\rightarrow  1$, indicating a completely aligned state. At
$\phi=-0.2$, the vertical red  line (dark gray) indicates that all values of
$0.2<S<0.4$ share the same energy, corresponding to an orbit of degenerate critical points of the free energy. The two dashed lines emanating from the origin
mark the boundary of the triangular region where the orientational distribution functions corresponding to equilibrium solutions do not have compact support, those outside  this region do.
\begin{figure}[th]
	\centering
	\includegraphics[width=.5\linewidth]{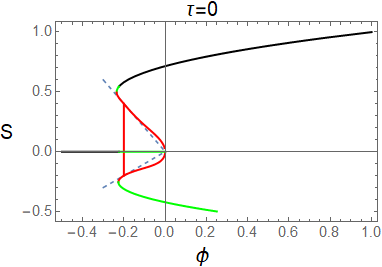}\caption{Equilibrium
		order parameter $S$ vs.\ $\phi$ for the solutions of the uniaxial self-consistent equations \eqref{eq:self_consistency} when $\tau=0$.  The black curve
		represents the stable branch of the bifurcation, the green curves (light gray) correspond
		to the metastable branches, and the red curves (dark gray) to the unstable branches.}%
	\label{fig-PRE}%
\end{figure}

\begin{figure}[ht]
	\begin{center}
		\begin{subfigure}{0.45\textwidth}
			\centering
			\includegraphics[width=\linewidth]{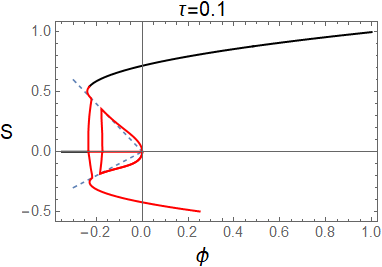}
			\caption{}
		\end{subfigure}
		\begin{subfigure}{.45\textwidth}
			\centering
			\includegraphics[width=\linewidth]{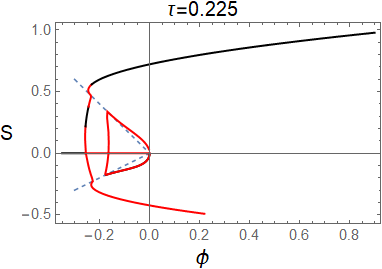}
			\caption{}
		\end{subfigure}
		\begin{subfigure}{.45\textwidth}
			\centering
			\includegraphics[width=\linewidth]{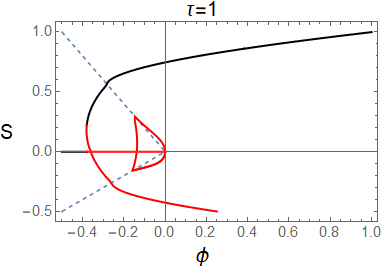}
			\caption{}
		\end{subfigure}
		\caption{Equilibrium order parameter $S$ vs. $\phi$ for the  solutions to the uniaxial self-consistency equations \eqref{eq:self_consistency}
			with $\tau =0.1, 0.225, 1$, respectively. The heavy black curves represent
			the globally stable branch, and the red curves (dark gray, in print) represent either metastable or unstable branches.}
		\label{fig-tau}
	\end{center}
\end{figure}

Now we consider the effect  of attractive interactions for some
representative values of $\tau>0$, which are shown in Fig.~\ref{fig-tau}. Due to
the increased complexity of the energy profiles, for simplicity, here we do not
distinguish metastable and unstable solution branches.  Interestingly, as $\tau$
increases, the vertical line at $\phi=-0.2$ in Fig.~\ref{fig-PRE} splits into two disjoint curves which move in opposite directions.
One of these curves borders an island, which eventually collapses at the
origin for $\tau\gg1$. The outer curve, which evolves in a
more interesting fashion, contains the essential phase transition information. The outer bifurcation curve admits noticeable kinks when it  nears the two dashed lines emanating from the origin; this appears to be evocative of some singularity of the free energy when regions of orientation space first become inaccessible.

More specifically, when $\tau$ is relatively small, say $\tau=0.1$, the
system undergoes a first order NI transition at $\phi_{NI}\approx-0.23$, with
$S_{NI}\approx0.54$, as shown in Fig.~\ref{fig-tau}(a).
As $\tau $ increases, both
$\phi_{NI}$ and $S_{NI}$ decrease.  Above a critical point $\tau
\approx0.22$, for example at $\tau =0.225$, there is a first order NI
transition occurring at $\phi_{NI}\approx-0.25$ with $S_{NI}\approx0.18$, as
shown in Fig.~\ref{fig-tau}(b), which is then followed by another
nematic-nematic transition at $\phi_{NN}\approx-0.24$ and $S_{NN}\approx0.53$.
The sudden drop of $S_{NI}$ at the critical $\tau $ occurs when the orientational distribution function corresponding to the order
parameter at the NI transition loses its compact support; the stable
nematic solution now lies in between the two dashed lines in
Fig.~\ref{fig-tau}(b). In this region, molecular orientation far from the director is allowed. When $\tau$ is large enough, the topology of the
bifurcation curves evolves further; for example, at $\tau=1$, the secondary NN transition disappears, and only the  primary NI transition occurs at
$\phi_{NI}\approx-0.38$ with $S_{NI}\approx0.24$. as shown in
Fig.~\ref{fig-tau}(c). As $\tau$ increases further, the outer open piece of
the bifurcation curve is stretched even more towards large negative values of
$\phi$.  At the critical value of $\tau\approx4.488$ the
leftmost point of the curve goes to $-\infty$. This means that below the
critical dimensionless temperature $1/\tau\rightarrow0.2228$, the nematic phase
can exist at arbitrarily low densities $\phi$, provided the temperature is sufficiently low. The value $1/\tau
=0.2228$ and the order parameter at the transition is fully
consistent with Maier-Saupe theory.

\begin{figure}[th]
	\centering
	\includegraphics[width=.5\linewidth]{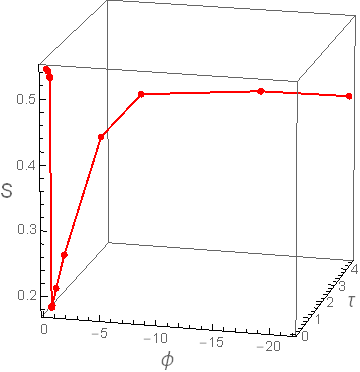}
	\caption{Equilibrium order
		parameter $S_{NI}$ and scaled density  $\phi_{NI}$ at the isotropic-nematic transition for the solutions to the uniaxial self-consistency equations \eqref{eq:self_consistency} plotted against the dimensionless reverse temperature $\tau$.}%
	\label{fig-3D}%
\end{figure}

The dependence of $S_{NI}$ and $\phi_{NI}$ on $\tau$ is shown in
Fig.~\ref{fig-3D}, which summarizes the dependence of the order parameter and
critical effective density at the NI transition on temperature.
As the
temperature decreases, the critical density and order parameter decrease
slowly initially. The attractive interactions stabilize the nematic phase, the stronger the attractive interactions, the less concentration is needed to form a nematic phase. As the temperature is further decreased, the order parameter
$S_{NI}$ suddenly drops at $\phi\approx0.22$, at which point the orientational
probability distribution function starts to lose its compact support and is positive everywhere in its orientational
space. As the temperature is decreased beyond this point, $S_{NI}$
increases  and saturates at $S_{NI}=0.43$, which agrees with the value predicted by
the Maier-Saupe theory.

A preliminary study shows that if the sign of the attractive interaction is changed, the special vertical line at $\phi=-0.2$ disappears, all stable nematic equilibrium distribution functions possess compact support, and transition occurs at a higher density, order parameter values at the transition increase as the absolute value of the interaction parameter gets larger.

A more careful and complete study of the model and phase diagram is currently
underway;  much remains to be explained. 

\section{Perspectives}
\label{sec:summary}

This paper was an attempt to combine in a unified model both attractive and repulsive sources of molecular interactions that can be responsible for the onset of nematic phases. Although this attempt was built on entirely new bases, it was not unprecedented. The need for combining these independent molecular nematogenic drives has been fueled by the weaknesses exhibited by both the Onsager and  Maier-Saupe theories when applied in solitude. The former is limited by the assumption on the diluteness of particles \cite{palffy-muhoray:missing}. The latter is sensitive to the assumption of spherical symmetry in the particle distribution, a small deviation from which can jeopardize the very existence of the ordering transition \cite{dejeu:role}.

Perhaps the first proposal for a combined theory of repulsive and attractive interactions as driving forces for the nematic ordering of molecules goes back to Gelbart and Baron~\cite{gelbart:generalized} (see also \cite{cotter:generalized}). This theory has often been referred to as the generalized van der Waals theory; it is computationally demanding and has been explicitly applied only for special repulsive potentials. These applications had however the merit of showing clearly that the anisotropy of the mean-field potential is mostly due to the interplay between the repulsive potential and the isotropic part of the attractive potential \cite{wulf:short-range,warner:interaction,eldredge:role}. For other similar models and generalized theories, the reader is referred to specialized reviews \cite{singh:phase,gelbart:van_der_Waals,gelbart:molecular}.

In a different, but concurrent vein, attractive dispersion forces have been combined with hard-core repulsion in a formal theory that has, as its crucial feature, a fourth-rank tensor, depending on the anisotropy of the interacting molecules \cite{sonnet:steric} (see also Sect.~1.4 of \cite{sonnet:dissipative}).

Finally, it is illuminating to recall the alternative approach followed by Luckhurst and Zannoni~\cite{luckhurst:why}. They envisaged short-range, repulsive interactions as responsible for the local organization of molecules in clusters, which in turn are subject to long-range, attractive interactions. This syncretic view posits that the molecular clusters aggregated by short-range interactions are not destroyed at the ordering transition, at which their long-range organization changes. According to this view, not molecules, but stable clusters would be subject to an effective pair potential of possible dispersion origins.

Our study was focused on the dense limit, where, in loose terms, molecules tend to form a few, highly pervasive clusters exhausting free volume. A quantitative description of this scenario is already highly problematic in the case of spherical particles with only steric interactions; various methods have been tried which may be worth combining in a unified setting \cite{fai:free}.

Even a greater challenge lies with extending any of these methods to anisomeric hard particles. Here we were contented with showing that attractive interactions do affect the ordering transition, but in the dense limit the scene is dominated by repulsive interactions, which tend to bind all particles in a few clusters. Studying the ways that can make this happen is the objective of future research.

\end{document}